\documentclass[aps,floats,twocolumn,epsf,prb,superscriptaddress,showpacs]{revtex4-1}
\usepackage{graphicx}
\usepackage{epstopdf}
\usepackage{amsmath}
\usepackage{amssymb}
\usepackage{booktabs}
\usepackage[colorlinks=true,urlcolor=blue,citecolor=blue,linkcolor=blue,breaklinks=true]{hyperref}
\usepackage{color}
\usepackage{bm}
\usepackage{mathpazo}
\usepackage{dsfont}
\usepackage[dvipsnames]{xcolor}
\usepackage{subcaption}
\usepackage{caption}
\usepackage[normalem]{ulem}

\usepackage{orcidlink} 

\definecolor{myRed}{RGB}{150, 0, 24}

 \DeclareCaptionJustification{links}{\raggedright}
\newcommand{\Ce}{Ce$_3$Bi$_4$Pd$_3$}

\begin{document}

\preprint{APS/123-QED}

\title{Weyl nodes in \Ce{} revealed by dynamical mean-field theory}

\author{Martin Bra\ss\,\orcidlink{0000-0002-4347-6987}}
\affiliation{Institute of Solid State Physics, TU Wien, 1040 Vienna, Austria}

\author{Jan M. Tomczak\,\orcidlink{0000-0003-1581-8799}}
\affiliation{Department of Physics, King’s College London, Strand, London WC2R 2LS, United Kingdom}
\affiliation{Institute of Solid State Physics, TU Wien, 1040 Vienna, Austria}

\author{Karsten Held\,\orcidlink{0000-0001-5984-8549}}
\affiliation{Institute of Solid State Physics, TU Wien, 1040 Vienna, Austria}

\date{\today}

\begin{abstract}
  Experimental studies have found unusual transport properties in \Ce{} which are potentially a consequence of the interplay between band-structure topology and electronic correlations. Based on these measurements, the existence of Weyl points in strongly renormalized, flat quasiparticle bands has been postulated. However, so far, there has been neither a direct spectroscopic observation of these, nor a calculation from first principles that would confirm their existence close to the Fermi energy. Here, we present density functional theory (DFT) and dynamical mean field theory (DMFT) calculations and study the low-energy excitations and their topological properties. We find that the Kondo effect promotes two out of the six angular momentum $J=5/2$  states, with the other four pushed  to higher energies. Further, we find  Weyl nodes close to the Fermi energy as previously suggested  for explaining the  observed giant spontaneous Hall effect in \Ce{}, as well as nodal lines.
\end{abstract}

\maketitle
 
\section{\label{sec:Introduction}Introduction}

Weyl-Kondo semimetals interface topology and strong correlation physics and exhibit low-energy excitations and transport properties different from weakly interacting materials.\cite{Dzsaber2017,xu2017,guo2017,guo2018,Dzsaber2021,lai2018} The Coulomb interaction between localized $f$-electrons and their hybridization with conduction bands results in flat bands and quasiparticles with very high effective masses and, accordingly, low renormalized velocities. These bands can cross each other at isolated points close to the Fermi energy giving rise to a linear dispersion relation of the quasiparticles which then behave like Weyl fermions\cite{herring1937,armitage2018}. Such Weyl points are monopoles of Berry curvature\cite{Volovik}, potentially providing significant contributions to the transverse electrical conductivity in Hall experiments, even without an external magnetic field.\cite{haldane2004}

\begin{figure}
    \centering
    \includegraphics[width=0.7\columnwidth]{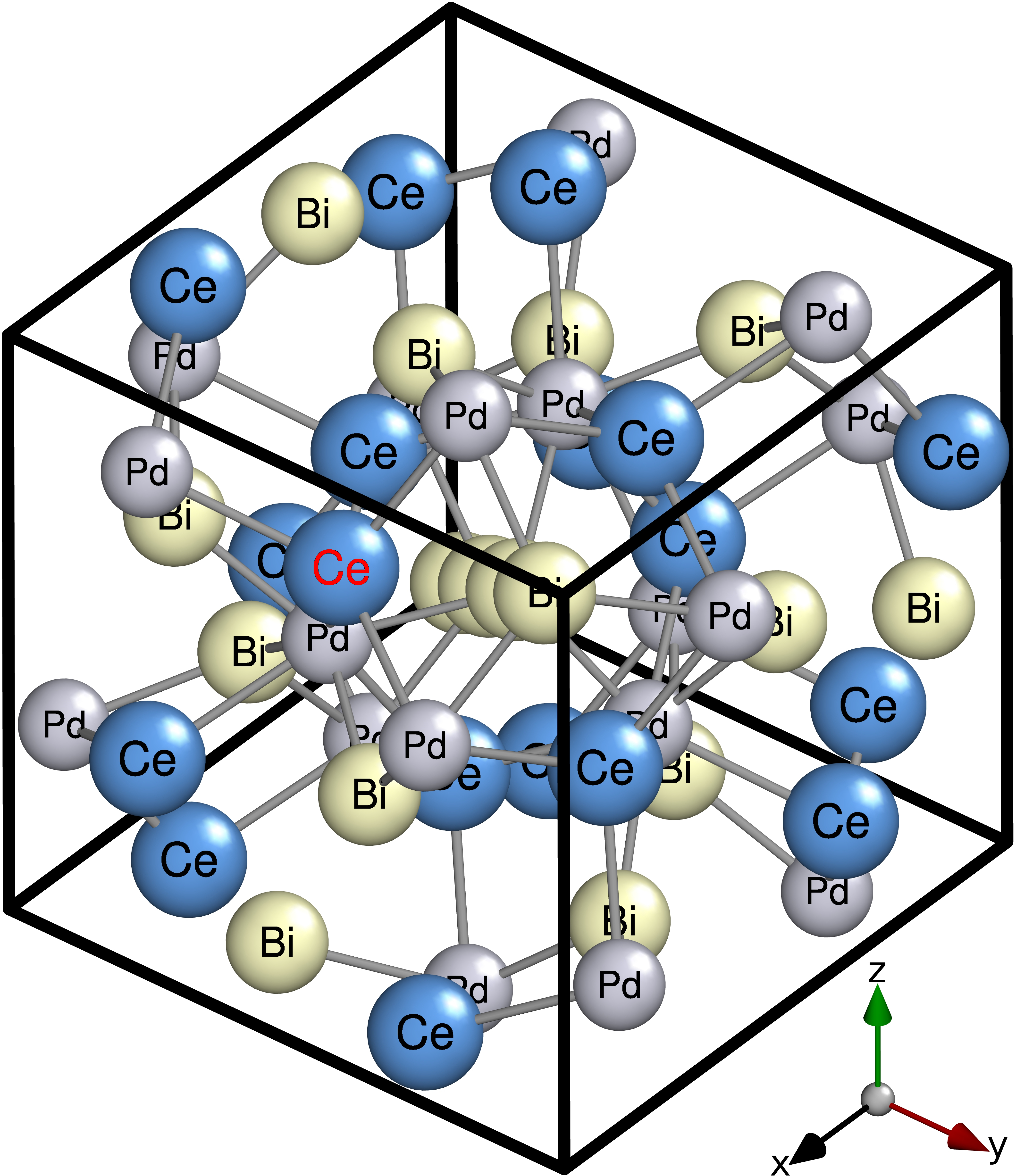}
    \caption{Crystal structure in the conventional unit cell of Ce$_3$Bi$_4$Pd$_3$. One of the equivalent Ce atom is highlighted for reference in Sec. \ref{sec:DFT}. }
    \label{fig:structure}
\end{figure}

One Weyl-Kondo semimetal candidate is the non-centrosymmetric compound \Ce{}\cite{Dzsaber2017} whose crystal structure\cite{hermes2008} is shown in Fig.~\ref{fig:structure}. The absence of inversion symmetry makes the existence of Weyl points possible and the nominally single, localized Ce-$4f$ valence electron gives rise to the Kondo effect at low temperatures\cite{Dzsaber2017}. When the system is cooled to the Kondo coherent regime, measurements of the specific heat show a cubic dependence on temperature. While usually this is a consequence of phonons, in the case of \Ce{}, the comparison to the reference material  La$_3$Bi$_4$Pt$_3$ indicates that the electronic contribution to the specific heat dominates the phononic one\cite{Dzsaber2017}. Hence, these observations have been attributed to the existence of Weyl fermions with quasiparticle velocities reduced by three orders of magnitude compared to weakly interacting metals\cite{Dzsaber2017}. 

Further evidence for this interpretation came from electrical conductivity measurements that showed a giant, spontaneous Hall effect.\cite{Dzsaber2021} Tilted Weyl points close to the Fermi edge could induce significant amounts of Berry curvature on the Fermi surface, which could explain the observed transverse electric current in the absence of magnetic fields\cite{lai2018,Dzsaber2021}.

However, so far, there has been neither a direct spectroscopic observation of these Weyl nodes in highly renormalized bands, nor a calculation from first principles that would confirm their existence in the vicinity of the Fermi energy. Here, we study \Ce{} by the combination\cite{kotliar2006electronic,held2007electronic}  of density functional theory (DFT) and dynamical mean-field theory (DMFT) that captures, both, the topological properties of the material, as well as the electronic correlations and Kondo physics. For the localized 4$f$-electrons of Ce, the local DMFT correlations can be expected to provide an accurate description, as long as we are not in the vicinity of an ordering instability with strong non-local correlations.

Our results are strikingly different from a previous DFT+DMFT study~\cite{PhysRevLett.124.166403} that did not discriminate between the self-energies of the different $J=5/2$ states ($J$: total angular momentum). As we will see below such a differentiation is essential to describe the Kondo effect in \Ce{} correctly. Further, in Ref.~\onlinecite{PhysRevLett.124.166403} topology has only been studied in the (effective one-particle) DFT not in the (interacting) DFT+DMFT electronic structure.

The outline of the paper is as follows:
In section \ref{sec:DFT} we describe the DFT electronic structure and discuss the symmetry properties of the local orbitals that will give rise to a Kondo resonance. In section \ref{sec:DMFT} we study the interacting band structure within DMFT. We derive an effective low-energy Hamiltonian in section \ref{sec:Hamiltonian}, which allows us to analyze its topology. We discover Weyl points from the renormalized quasiparticle bands in section \ref{sec:results} and nodal lines in section \ref{sec:nodal_lines}. A summary and discussion of our results can be found in section \ref{sec:Conclusion}.

\section{\label{sec:DFT}Tight-binding model from first principles}

To study how electronic correlations of the localized Ce-4$f$ orbitals influence the topological properties, we need an accurate and material-realistic model that captures the hybridization between these. As a starting point, we perform a self-consistent density functional theory (DFT) calculation using the full potential local orbital (\textsc{FPLO}) code \cite{koepernik1999}. We employ a dense $(6\times 6\times 6)$ $k$-mesh and the Perdew-Wang exchange-correlation potential \cite{perdew1992} to determine the electronic band structure. After the calculation is converged, we take the resulting ground state density as starting point for another self-consistent computation, where we increase the $k$-mesh to $(12\times 12\times 12)$. This computation converges immediately, thereby confirming that the initial $k$-mesh is sufficient. All calculations are fully relativistic, which especially includes spin-orbit coupling (SOC). 

\begin{figure*}
    \centering
    \begin{subfigure}{0.49\textwidth}
    \includegraphics[width=\textwidth]{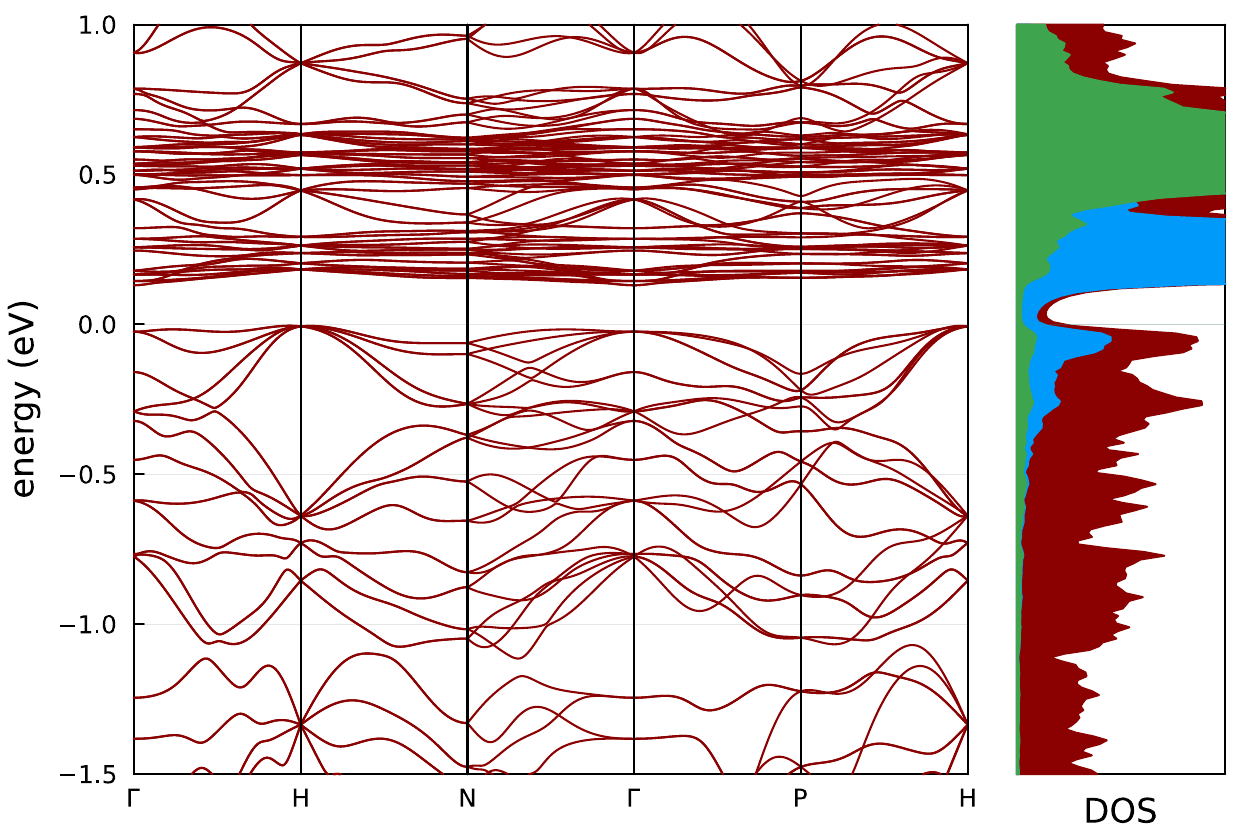}
    \caption{\label{fig:DFT_bands_4f_valence}}
    \end{subfigure}
    \begin{subfigure}{0.49\textwidth}
    \includegraphics[width=\textwidth]{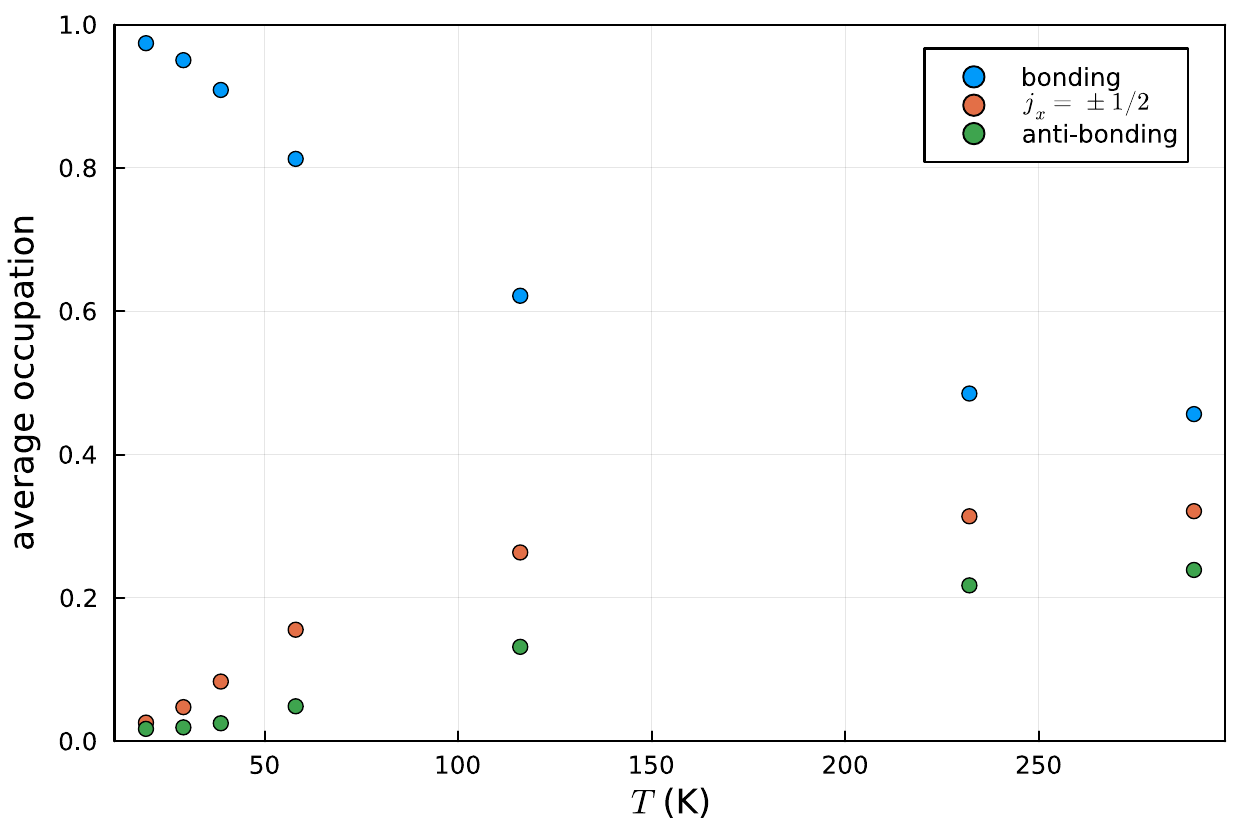}        
    \caption{\label{fig:occupations_4f}}
    \end{subfigure}

    \caption{\label{fig:DFT_bands} (a) Electronic band structure calculated within DFT. In the primitive basis the coordinates of the high symmetry points are: $H=(1/2,-1/2,1/2),\,\, P=(1/4,1/4,1/4),\,\, \Gamma=(0,0,0),\,\, N=(0,0,1/2)$. The Ce-4$f$ orbitals hybridize with the conduction bands, which leads to a band gap. The right panel shows the total density of states (red) as well as the partial densities for the Ce-4$f_{5/2}$ (blue) and Ce-4$f_{7/2}$ (green) orbitals. (b) Average occupation of the Ce-4$f_{5/2}$ orbitals as function of temperature within DMFT.}
\end{figure*}

In Fig.~\ref{fig:DFT_bands_4f_valence} we show the corresponding band structure. We obtain an insulating system in DFT, because hybridization between Ce-4$f$ orbitals and conduction bands compresses the latter and opens a gap. By looking at the density of states (DOS) in the right panel of Fig.~\ref{fig:DFT_bands_4f_valence}, we can see that the Ce-4$f$ orbitals in Ce$_3$Bi$_4$Pd$_3$ have a contribution close to the Fermi energy. If electronic correlations are later included, these will give rise to a narrow Kondo-peak at the Fermi energy. 
However, in order to describe this Kondo physics, we need to treat the electronic correlations more accurately than it is possible within DFT. 

To capture the low-energy degrees of freedom of the system, we perform a Wannierization of the bands. As there are many entangled bands around the Fermi energy, we here project them onto the $4f$ orbitals of the six Ce atoms, the $4d$ orbitals of the six Pd atoms and the $6p$ orbitals of the eight Bi atoms in the unit cell. This yields altogether 192 Wannier-orbitals out of which $6\times 14$ correspond to the Ce-$4f$. For the latter, correlations need to be treated beyond a static mean-field approximation. This large number of correlated orbitals makes computations very demanding. However, we can simplify the problem, by observing that SOC nicely separates $4f_{5/2}$ and $4f_{7/2}$ orbitals in energy as can be seen in the DOS from Fig.~\ref{fig:DFT_bands_4f_valence}. The $4f_{7/2}$ orbitals are farther above the Fermi energy, such that only the $4f_{5/2}$ orbitals will be occupied at low temperatures. This energy separation allows us to only treat the correlations of the $4f_{5/2}$ orbitals on the level of dynamical mean-field theory (DMFT) and the Coulomb repulsion between $4f_{7/2}$ and $4f_{5/2}$ orbitals on a static mean-field level.

By analyzing the symmetry properties of the $4f_{5/2}$ orbitals, we can identify which of them can interact with each other and which cannot. For this, we compute the stabilizer group of the Ce atoms, i.e. the sub-group of space-group $I\bar{4}d$ (220) which leaves the position of a Ce atom invariant. In the present case the stabilizer is generated by a single fourfold screw roto-inversion axis, that is a fourfold rotation, followed by inversion and translation by a fractional lattice vector. In other words, the stabilizer is isomorphic to $S_4$. 

Hence, the Ce-$4f$ orbitals form a representation of this group. However, as SOC is not negligible, the group of relevance here is the double cover of $S_4$ for which the $4f_{5/2}$ orbitals form a representation which we can decompose into irreducible representations (irreps) using the calculus of characters. Since $S_4$ is abelian, all irreps are one dimensional. Additionally, due to time reversal symmetry, for every irrep contained in the space spanned by the $4f_{5/2}$ orbitals its dual representation must be contained as well such that these are degenerate in the local crystal field. Thus the latter will split the $4f_{5/2}$ manifold into three twofold degenerate states. A direct calculation reveals that the decomposition reads
\begin{equation}
        \Gamma_7\oplus \Gamma_7^\star\oplus\Gamma_6^{\oplus 2}\oplus\left(\Gamma_6^\star\right)^{\oplus 2}
\end{equation}
Here we used the notation for irreps from Cracknell\cite{cracknell1979}, a star denotes the dual representation, $\oplus$ the direct sum of vector spaces and the exponent $\oplus 2$ indicates that an irrep occurs twice in the decomposition.

To get a better intuition of what these irreps are, we exemplarily pick the Ce atom highlighted in Fig.~\ref{fig:structure} in red. For this atom, the $S_4$ axis is parallel to the x-axis. Then we can choose the spin-quantization axis to be the x-axis as well and observe that the $j_x=1/2$ spans the $\Gamma_7$ irrep and $j_x=-1/2$ spans its dual. Furthermore, both $j_x=3/2$ and $j_x=-5/2$ transform as $\Gamma_6$. Therefore, these orbitals can interact and will hybridize due to the local crystal field. The same holds for $j_x=-3/2$ and $j_x=5/2$ which both transform as the dual irrep $\Gamma_6^\star$.

Due to hybridization these states will form a bonding and an anti-bonding orbital, whose onsite energies can be obtained from our Wannier-Hamiltonian. The bonding orbital is lower in energy than the $j_x=1/2$ orbital which in turn is lower in energy than the anti-bonding orbital.

To speed up the following DMFT calculation we take advantage of the above observations. If we transform the Wannier-orbitals into the basis of bonding, anti-bonding and $j_x=\pm 1/2$ orbital, then locally the Ce-$4f$ part of the Hamiltonian is diagonal. During the DMFT cycle the $j_x=1/2$ orbital will not obtain off-diagonal terms, since it belongs to an irrep which occurs only once in the decomposition of the $4f_{5/2}$ manifold. For the bonding and the anti-bonding orbital off-diagonal elements might be generated, since these orbitals can hybridize further, but they are small and hence we will neglect them in the DMFT calculation. This gives a considerable performance boost, which allows us to go to low temperatures.

\section{\label{sec:DMFT}Calculations with DMFT}
With the Wannier Hamiltonian from the previous section we have a model at hand which is derived from first principles and can be used as starting point for a DMFT calculation. Here, we use the quantum Monte Carlo (QMC) continuous time hybridization expansion (CT-Hyb) as impurity solver using the \textsc{w2dynamics} implementation\cite{WALLERBERGER2019388}. The electronic interaction is modeled by a density-density interaction with Coulomb repulsion parameter $U = 6$ eV, similar to values used previously for various Ce$_3$A$_4$M$_3$ compounds\cite{jmt_CBP_arxiv,PhysRevLett.124.166403,Xu2022,PhysRevB.108.245125}. Double counting is taken into account as described by Anisimov et al.\cite{Anisimov_1997} Since Ce is in a dominantely 4$f^1$ configuration, we neglect Hund's exchange for the sake of reaching lower temperatures in DMFT. This is justified as long as we are not looking into the multiplet splitting of the upper Hubbard bands. Using Pulay mixing\cite{PULAY1980393} (DIIS), DMFT converges after 60 iterations.

\begin{figure*}
    \centering
   \begin{subfigure}{0.49\textwidth}
        \includegraphics[width=\textwidth]{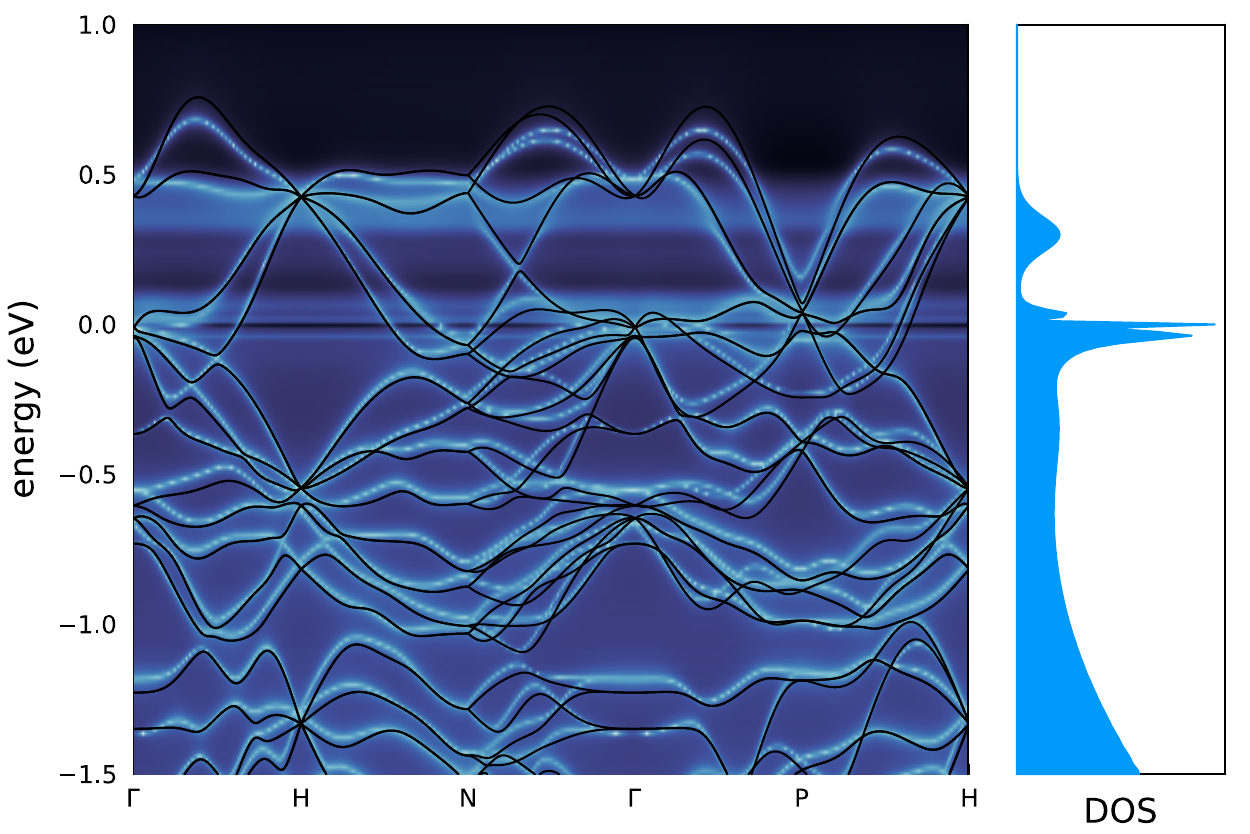}
        \caption{\label{fig:DMFT_bands}}
    \end{subfigure}
    \begin{subfigure}{0.49\textwidth}
        \includegraphics[width=\textwidth]{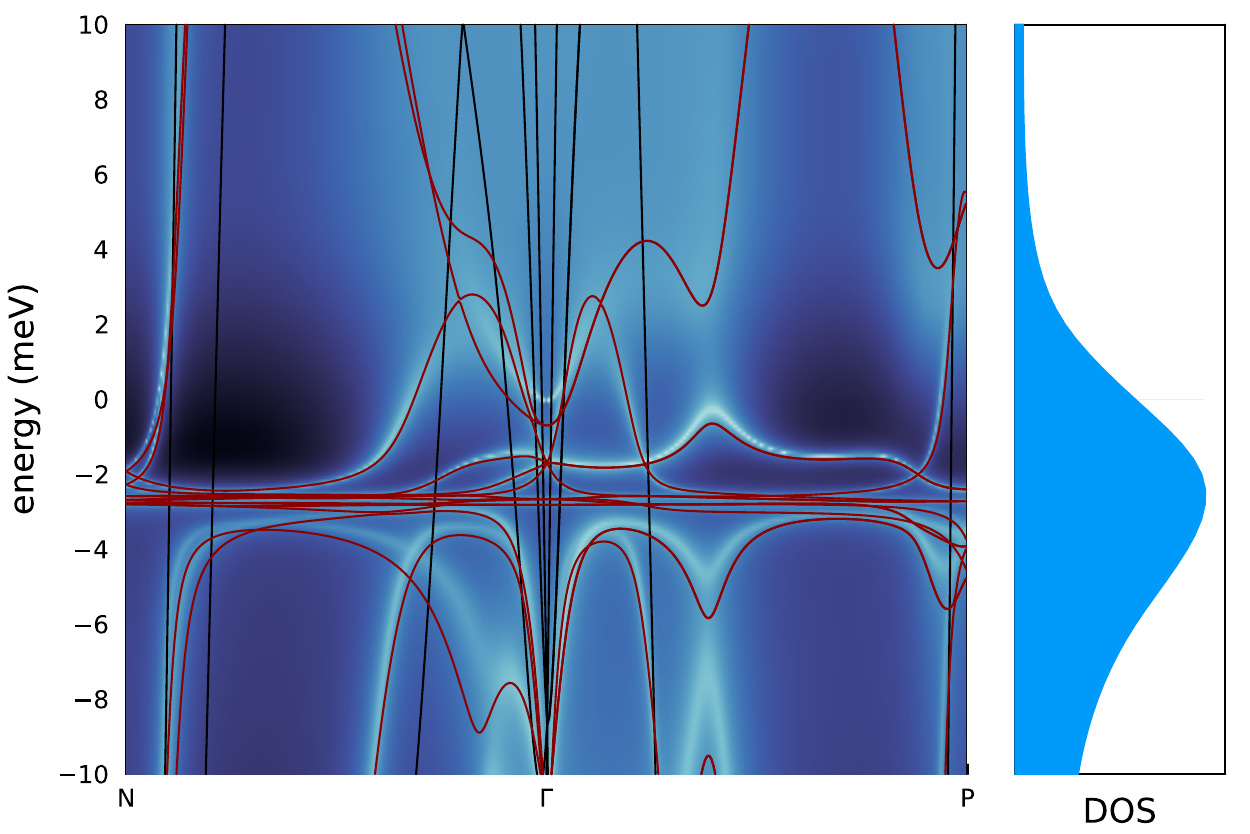}
        \caption{\label{fig:DMFTzoom}}
    \end{subfigure}
    
    \caption{\label{fig:DMFT} (a) Momentum-resolved spectral function from DMFT at $T=29$K (false color; intensity on logarithmic scale for better visibility). The black curve is the DFT band structure where the Ce-4$f$ orbitals are treated in the open-core approximation. The rightmost panel of (a) shows the partial density of states of the bonding Ce-4$f_{5/2}$ orbital with an emerging Kondo resonance. (b) Zoom-in along N-$\Gamma$-P. Close to the Fermi energy the quasiparticle Hamiltonian from Eq.~(\ref{eq:effectiveH}) (red curve)  agrees much better with DMFT than the open core DFT bands (black).}
\end{figure*}

First, we study the average occupation numbers of the Ce-$4f_{5/2}$ orbitals as a function of temperature. For a converged DMFT calculation they can be directly obtained from the average occupations of the impurity. They are shown in Fig. \ref{fig:occupations_4f}. While at room temperature all six orbitals have comparable occupations, the Kondo effect changes this. Specifically, the lower the  temperature becomes, the more depleted are the $j_x=\pm 1/2$ and anti-bonding orbitals. The two bonding orbitals however, tend towards being occupied by one electron in total.  Hence, at low temperatures solely these 4$f$ orbitals with the lowest local crystal field potential are occupied. This state has a local magnetic moment, and forms a Kondo resonance through hybridization with the conduction electrons.

In order to resolve this Kondo resonance, we need to analytically continue the imaginary time QMC data obtained within DMFT. This is performed using the maximum entropy method as implemented in $\Omega$MaxEnt\cite{OmegaMaxent}. We analytically continue both the local Green's function as well as the self-energy. The imaginary part of the local DMFT Green's function gives us the partial density of states for the Ce-$4f$ bonding orbital as shown in the right panel of Fig. \ref{fig:DMFT_bands}. There, we can see a Kondo resonance at the Fermi energy $\epsilon_F=0\,$eV. As expected already from the occupations, it has almost solely contributions from the bonding orbital.

The k-resolved spectral function in the left panel of Fig. \ref{fig:DMFT_bands} is obtained as the imaginary part of the interacting lattice Green's function with the analytically continued DMFT self-energy. Since the DMFT self-consistency loop started from the non-interacting band structure in Fig. \ref{fig:DFT_bands_4f_valence}, which includes hybridization between Ce-$4f$ and conduction bands, we can observe, how interactions influence the excitation spectrum: At low temperatures a Kondo resonance emerges in the immediate vicinity of the Fermi level.

Seemingly similar dispersions can be observed if we perform a DFT calculation where the Ce-4$f$ orbitals are treated in the open-core\cite{} approximation. The bands from this calculation are shown by the black lines in Fig. \ref{fig:DMFT_bands}, and are in good agreement with previous results.\cite{PhysRevLett.124.166403} 

Both calculations yield similar features if we are sufficiently away from the Fermi energy. However, in DMFT,   flat renormalized quasiparticle bands  emerge in the spectral function close to the Fermi energy due to the Kondo effect and show up as essentially horizontal lines in Fig. \ref{fig:DMFT_bands}. The zoom in Fig. \ref{fig:DMFTzoom} shows them more clearly and reveals that they are dispersive on an meV scale. At $T=29$K they are not fully coherent throughout the whole Brillouin zone, as can be seen by a pronounced smearing.
However, at temperatures sufficiently smaller than the Kondo temperature  this smearing  (the imaginary part of the self energy) is expected to go away. The flat quasiparticle bands will become sharp and, potentially, important for transport. Hence, in the next section we study their topological properties.

If we zoom in on the Fermi edge in Fig. \ref{fig:DMFTzoom} we can see that the open core DFT calculation does not capture the  renormalized flat bands at all, which shows that they are a consequence of electronic correlations. Energetically the maximum of the Kondo peak is approximately $\omega_0=2.4$ meV below the Fermi energy, but with $\epsilon_F$ still within the width of the Kondo resonance, as can be seen from the right panel of Fig \ref{fig:DMFTzoom}. Upon reducing temperature and thus the smearing, the Kondo peak will become sharper; and  we expect an upshift of the Kondo resonance since it necessarily forms around $\epsilon_F$.


In the spectral function and also in the effective low energy Hamiltonian discussed in the next section, we do not observe a Kondo insulating gap throughout the full Brillouin zone suggesting semi-metallic behavior down to lowest temperatures. We note that the related (isoelectronic) compounds Ce$_3$Bi$_4$Pt$_3$ and Ce$_3$Sb$_4$Pt$_3$ are, instead, prototypical Kondo insulators \cite{PhysRevB.42.6842,PhysRevB.58.16057,NGCS}. Owing to their larger Ce-4$f$ to conduction states hybridization,\cite{jmt_radialKI,PhysRevB.106.L161105} the hybridization gap, while renormalized, remains finite when including correlation effects.

\section{\label{sec:Hamiltonian}Effective low energy Hamiltonian}
In order to search for Weyl-points in the quasiparticle bands, we need an effective single particle Hamiltonian that captures the low energy physics of the system. From the DMFT calculation and the analytically continued self-energy $\Sigma(\omega)$ we can determine the momentum dependent Green's function
\begin{equation}
    \label{eq:GreensFunction}
    G(\omega,\mathbf{k}) = \left(\omega - H_\mathbf{k} - \Sigma(\omega)-\Delta_{\mathrm DC} +\mu\right)^{-1}
\end{equation}
Here $H_\mathbf{k}$ denotes the Bloch-Hamiltonian of the non-interacting Wannier-model derived from DFT. The self-energy $\Sigma$ and the double counting correction $\Delta_{\mathrm DC}$ are diagonal matrices which are non-zero only for the bonding, anti\-bonding and $j_x=\pm 1/2$ Ce-$4f_{5/2}$ orbitals. $\mu$ is the chemical potential from DMFT. 

We expand the Green's function around the maximum of the Kondo resonance $\omega_0$ 
\begin{equation}
\label{eq:approxGF}
    G(\omega,\mathbf{k})\! \approx\!\frac{Z}{\omega\! -\! Z^{\frac{1}{2}}\left(H_\mathbf{k}\!+\!\Sigma(\omega_0)\!-\!\omega_0\frac{\partial\Sigma(\omega_0)}{\partial\omega}\!+\!\Delta_{\mathrm DC}\! -\!\mu\right)Z^{\frac{1}{2}}}
\end{equation}
where we introduced the quasiparticle renormalization $Z = \left(1-\frac{\partial\Sigma(\omega_0)}{\partial\omega}\right)^{-1}$ which can be obtained via numeric differentiation. At the peak maximum $\omega_0$ we find that the imaginary part of the self-energy is minimal and thus $Z$ is real valued.

Eq. (\ref{eq:approxGF}) resembles the Green's function of an effective, renormalized quasiparticle Hamiltonian
\begin{equation}
\label{eq:effectiveH}
    H_\mathrm{qp}(\mathbf{k}) = Z^{\frac{1}{2}}\left(H_\mathbf{k}+\Sigma(\omega_0)-\omega_0\frac{\partial\Sigma(\omega_0)}{\partial\omega}+\Delta_{\mathrm DC} -\mu\right)Z^{\frac{1}{2}}
\end{equation}
Taking the square root of $Z$ ensures that Eq. (\ref{eq:effectiveH}) describes a hermitian operator. We compare its band structure to the DMFT spectral function in Fig. \ref{fig:DMFTzoom} (red lines), where we can see that both agree quite well in a region of a few meV around the Fermi energy. The quasiparticle Hamiltonian captures the low energy excitations due to electronic correlations which the open core DFT calculation (black lines) cannot describe. Therefore, $H_\mathrm{qp}$ is a good starting point to study whether \Ce{} exhibits Weyl-points in the vicinity of the Fermi energy, which we pursue in the next section. 

The reason why the quasiparticle Hamiltonian describes the low energy excitations better than open core DFT lies in the quasiparticle renormalization $Z$ which in the present case is small thereby strongly renormalizing the width of the Ce-$4f$ bands leading to almost flat bands. This is also an important difference compared to the topological Hamiltonian\cite{Wang2013} $H_\mathrm{topo}(\mathbf{k})=G^{-1}(\omega=0,\mathbf{k})$ which is the inverse of the Green's function evaluated at zero frequency. It differs from $H_\mathrm{qp}$ by the energy shift $\omega_0\partial_\omega\Sigma(\omega_0)$ and the scaling $Z^{\frac{1}{2}}$. The quasiparticle renormalization is essential to obtain the flat quasiparticle bands close to the Fermi energy which are not present in the topological Hamiltonian.

\section{\label{sec:results}Weyl points in quasiparticle bands}
Now we search for Weyl-points in $H_\mathrm{qp}$. Since  the Wannierization led to a tight-binding Hamiltonian with 192 spin-orbitals as described in the previous sections, this search is computationally demanding. Therefore, we employ our recently described algorithm\cite{brass2024} which traces the Berry curvature vector-field to its sinks and sources via solving an ordinary differential equation.

\begin{figure}
    \centering
    \includegraphics[width=\columnwidth]{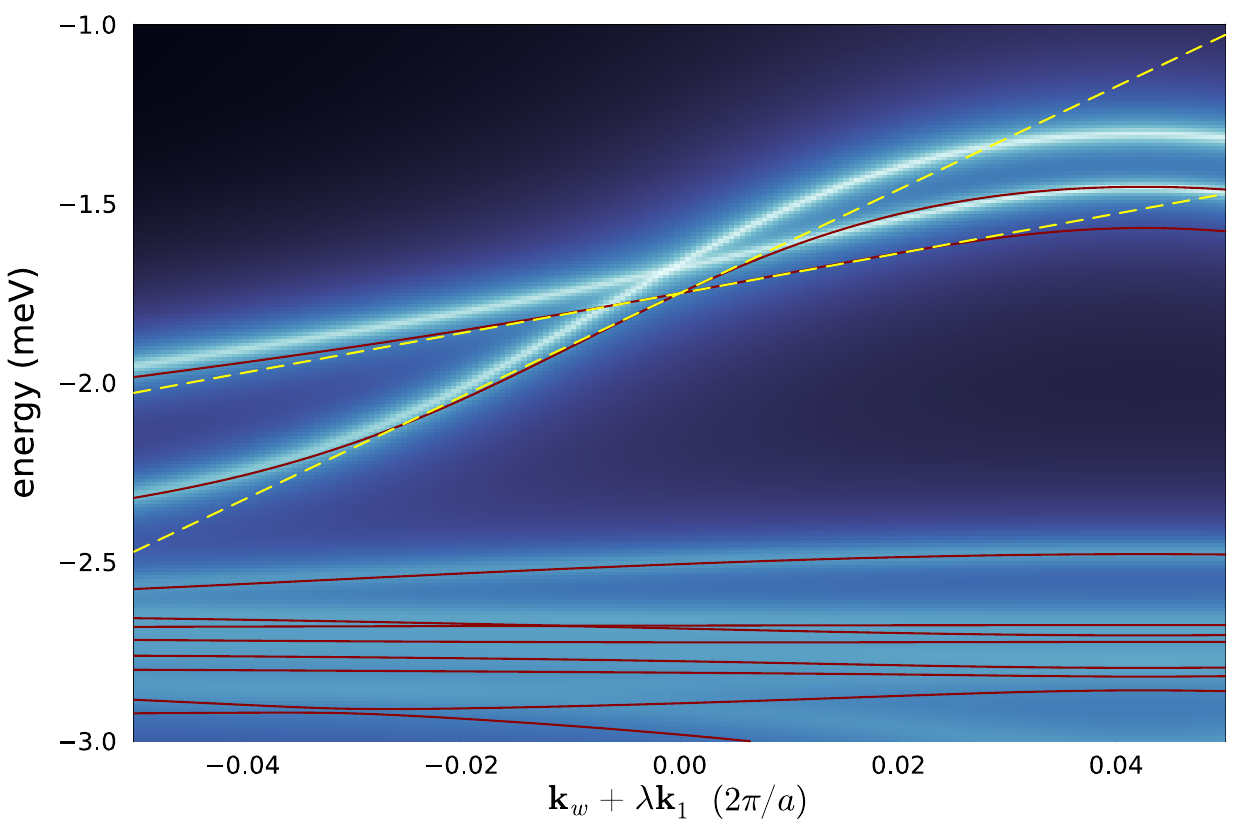}    
    \caption{\label{fig:WeylPoint} At $T=29$K the quasiparticle renormalized Hamiltonian $H_\mathrm{qp}$ from Eq. (\ref{eq:effectiveH}) has a Weyl point a few meV below the Fermi energy (red lines). These band crossings can also be found in the momentum resolved spectral function with the DMFT self-energy (logarithmic color-scale). Since the tangents of both crossing bands (yellow dashed lines) have positive slope, we can infer that this is a type-II Weyl point. The path in momentum space is chosen as $\mathbf{k}_W+\lambda \mathbf{k}_1$ where $\mathbf{k}_W$ is the momentum of the second Weyl point from Table~\ref{tab:weylpoints}, $\mathbf{k}_1$ is the first primitive basis vector, and $\lambda\in\left[-0.05,0.05\right]$. }
\end{figure}

We find nine symmetrically in-equivalent band crossings in the vicinity of $\epsilon_F$, each of which sits at general momenta away from high symmetry lines. We confirm that these are Weyl-points by calculating the corresponding Chern numbers numerically. Due to time reversal and the 24 point group symmetries,  each Weyl node is 48-fold degenerate, i.e. it belongs to a set of 48 nodes related by symmetry. As half of the point group symmetries have negative determinants, each of these sets can be split into two halves of nodes having opposite Chern numbers. Hence, the Nielsen–Ninomiya theorem which enforces a total of zero Chern numbers \cite{nielsen1981,nielsen1981a} is fulfilled.

Four sets of Weyl points are a few meV below, five a few meV above the Fermi edge. From the eigenstates of $H_\mathrm{qp}$ we can infer, that the former four sets belong to bands whose character is around 90 to 99$\%$ of the bonding $4f_{5/2}$ orbital. The latter five sets above $\epsilon_F$ have predominantly $j_x=\pm 1/2$ character. We list them in table \ref{tab:weylpoints} and exemplarily show the second Weyl-point in Fig.~\ref{fig:WeylPoint}. By comparing the band structure of the effective quasiparticle Hamiltonian $H_\mathrm{qp}$ (black lines) to the momentum resolved spectral function from DMFT (color-scale), we can see that this band-crossing is present in both of them.

\begin{table}
    \centering
    \begin{tabular}{|c|c|c|}
    \hline location ($2\pi/a$) & energy (meV) &  type\\\hline
    (0.584,\, 0.082,\, -0.225)   & -1.26 &  II\\
    \textcolor{myRed}{(-0.568,\, -0.394,\, -0.691)} & \textcolor{myRed}{-1.75}  & \textcolor{myRed}{II}\\
    (0.559,\, 0,\, -0.289)     & -2.14 & I\\
    (-0.354,\, 0,\, -0.496)   & -2.42 & II\\
    (-0.371,\, -0.728,\, -0.495)   &     7.27  & II\\
    (-0.348,\, 0.398,\, 0.181) & 21.08  & II \\
    (0.262,\, 0,\, -0.493)   & 26.18  & II\\
    (-0.251,\, -0.559,\, -0.006)  &      26.35  & II\\
    (0.574,\, 0,\, 0.228)   &  26.41 &  II\\
    \hline
    \end{tabular}
    \caption{Weyl points of the quasiparticle Hamiltonian. Their locations are listed in Cartesian coordinates rounded to three digits. We present only those Weyl points that belong to bands closest to the Fermi energy, but others exist, too. Due to symmetry, each point is 48-fold degenerate. The dispersion around the second Weyl point (marked in red) is shown in Fig.~\ref{fig:WeylPoint}.} 
    \label{tab:weylpoints}
\end{table}

With energies differing only on a sub-meV scale compared to the DMFT spectral function, $H_\mathrm{qp}$ is a sufficiently good approximation to detect Weyl-points in momentum space. The reason for small energetic offsets is that $H_\mathrm{qp}$ is obtained from an expansion of the DMFT Green's function around the maximum of the Kondo peak. If we go away from this peak, we must expect some deviations in energy. However, as long as these deviations do not become large enough to drastically move the Weyl points or create or annihilate them in pairs, $H_\mathrm{qp}$ can be used to search for topological band crossings and compare them to the DMFT spectral function. At least this is the case for the four Weyl nodes below the Fermi edge; those five above are too far away in energy from the maximum of the expansion for a meaningful comparison of the results from $H_\mathrm{qp}$ to the DMFT spectral function. Furthermore, as discussed earlier, they have predominantly $j_x=\pm 1/2$ character. Hence, to study these it would be better to obtain $H_\mathrm{qp}$ from an expansion around the peak energies of these orbitals. 

The plot reveals the type-II nature of the Weyl node, which can be inferred from the slopes of the crossing bands having the same sign\cite{soluyanov2015,armitage2018}. This discovery of a type-II Weyl point supports the interpretation of the spontaneous Hall effect seen in experiment. Furthermore, its type-II nature impacts the temperature dependence of specific heat capacity $c_V$. While a type-I Weyl node has a point like Fermi surface and therefore leads to a cubic temperature dependence of $c_V$,\cite{Dzsaber2017,lai2018} a type-II node is embedded in a finite Fermi surface\cite{soluyanov2015} such that the specific heat has terms linear and cubic in temperature. This reasoning can be confirmed by calculating $c_V$ from the quasi particle Hamiltonian with the chemical potential fixed to the energy of the first Weyl point in table \ref{tab:weylpoints} and fitting $c_V = aT+bT^3$ to these data. From the cubic term we can calculate the quasi particle velocity\cite{Dzsaber2017,lai2018} via $v^\star=\sqrt[3]{7\pi^2k_B/(30b)}$ yielding approximately $245\,\mathrm{m/s}$. Similar to the experimental results\cite{Dzsaber2017} this is three orders of magnitude smaller than for usual, weakly interacting metals because of the flatness of the quasi particle bands. Hence, our calculations support the experimental findings qualitatively. Quantitatively however, our result is by a factor of 3.6 smaller than experiment, likely due to the fact that here we used the quasi particle Hamiltonian to determine $c_V$. For a quantitative result the specific heat should be calculated within DMFT for temperatures between two and ten Kelvin\cite{Dzsaber2017}.

To test the stability of our findings we apply the procedure described above to the quasiparticle Hamiltonian (\ref{eq:effectiveH}) obtained from the self energy of the converged DMFT calculation and from the self energy averaged over the last five DMFT iterations. It turns out that although the height of the Kondo resonance fluctuates slightly, the momenta of the Weyl points are almost unaffected. This reflects the fact that Weyl points are topological properties of the band structure and hence small perturbations of the system do not remove them unless the perturbation is large enough to annihilate them in pairs, which is not the case here.

On the other hand, the topological character of the Weyl nodes raises the question whether these nodes are already present at the DFT level, but adiabatically moved to different positions in energy and momentum. In the case of the DFT calculation including the Ce-$4f$ orbitals, there is a band gap (which closes within DMFT). Hence, the DFT Hamiltonian and the quasiparticle Hamiltonian $H_{qp}$ are not adiabatically connected. Consequently, we cannot expect that they share topological features and indeed we did not find Weyl-points close to the Fermi edge in the DFT Hamiltonian. Thus, the low-energy Weyl points are correlation-induced. 

Due to the apparent similarity of the DFT bandstructure where Ce-$4f$ orbitals are treated in the open-core approximation and the DMFT spectral function in Fig. \ref{fig:DMFT_bands}, we revisit the former and perform a search for Weyl-points with the same algorithm\cite{brass2024} as above. We detect multiple Weyl points as listed in table \ref{tab:weylpoints_DFT}, which also includes Weyl points that had not been found in an earlier study\cite{Dzsaber2021}.

\begin{table}
    \centering
    \begin{tabular}{|c|c|c|}
    \hline location ($2\pi/a$) & energy (meV)&  type \\\hline
    (-0.77,\, -0.862,\, -0.656) & -18.1  & II\\
    (-0.418,\, -0.566,\, -0.348)  & -43.0  & II\\
    (0.527,\, -0.142,\, 0.274)  & -118 &  II\\
    (0.546,\, 0,\, -0.231)  &  -132 &  II\\
    (0.07,\, -0.438,\, -0.324) & -157 &  II\\
    (-0.395,\, 0,\, 0.425)  &  -161 &  II\\
    (0.182,\, 0,\, 0.464)   &  -180.7  & II\\
    (-0.145,\, -0.426,\, 0.253) & -181.4  & II\\
    (-0.587,\, -0.769,\, -0.802)  & -196  & II\\
    (0.109,\, 0,\, 0.39)  &    -208 &  I\\
    (0.494,\, 0.0,\, -0.248) &   247 & II\\
    (0.705,\, 0,\, -0.129) &   -310 &  II\\
    (0.114,\, 0.571,\, -0.223) &   318 & I\\ 
    (-0.779,\, -0.782,\, -0.992) & -420 & II\\
    \hline
    \end{tabular}
    \caption{Weyl points from DFT where Ce-$4f$ orbitals are treated in the open-core approximation. Their locations are listed in Cartesian coordinates rounded to three digits.}
    \label{tab:weylpoints_DFT}
\end{table}
Some of these also appear in the DMFT spectral function at almost the same momentum and energy. Others may change their position or be pairwise annihilated. However, none of them are as close to the Fermi edge as the Weyl nodes from the renormalized quasiparticle bands.  

Beyond this disparity, there is an important physical distinction between the two settings.
The open-core DFT calculation, used previously to advocate a topological state in Ce$_3$Bi$_4$Pd$_3$\cite{PhysRevLett.124.166403}, mimics effectively localized f-states, corresponding to the {\it high-temperature} local-moment regime. The signatures of the topological state, however, are observed at very {\it low-temperatures}, where an additional Kondo resonance emerges due to many-body effects. This yields the flat renormalized quasiparticle bands close to the Fermi edge which are not present in DFT and are essential for the existence of the discussed Weyl nodes. The apparent similarity of the conduction band dispersions thus does not justify a meaningful analysis of topology on the basis of DFT.

\section{\label{sec:nodal_lines}Nodal lines in quasiparticle bands}
\begin{figure*}
    \centering
    \begin{subfigure}{0.49\textwidth}
    \includegraphics[width=\textwidth]{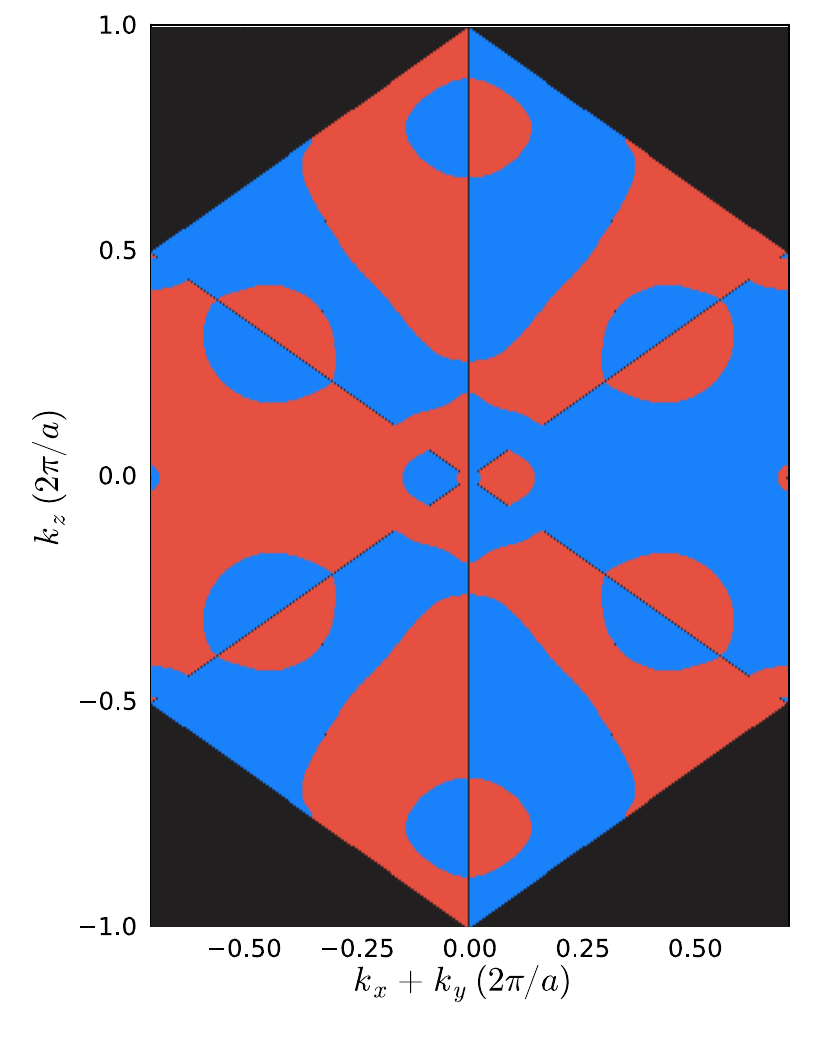}        
    \caption{\label{fig:mirror}}
    \end{subfigure}
    \begin{subfigure}{0.49\textwidth}
    \includegraphics[width=\textwidth]{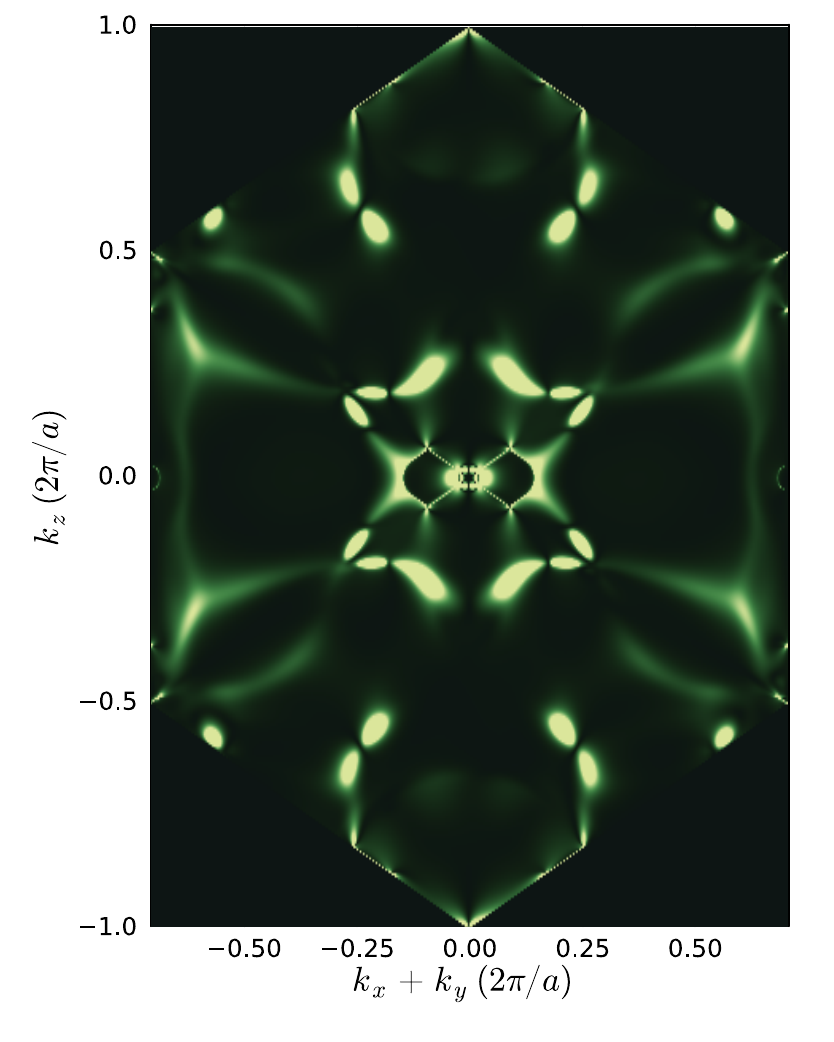}
    \caption{\label{fig:berry}}
    \end{subfigure}
    \caption{\label{fig:mirror_berry} (a) Eigenvalues of glide mirror symmetry in the plane spanned by $k_z$ and $k_x+k_y$ for a renormalized quasiparticle band: red and blue denote $\pm i$ respectively. The boundaries of areas of different mirror eigenvalues are nodal lines. (b) Magnitude (norm) of Berry curvature in the same plane. By comparison with (a) we see that Berry curvature is large near some of the nodal lines, but also at other points.}
\end{figure*}

Besides Weyl nodes other types of topological band crossings are possible. One example is the nodal line where bands are degenerate on a curve in momentum space. A previous study\cite{PhysRevLett.124.166403} found nodal lines in their Bloch Hamiltonian from a DFT calculation with Ce-$4f$ orbitals treated in the open-core approximation as well as in the {\it topological} Hamiltonian\cite{Wang2013} extracted from their DMFT calculation. Here, we study whether the renormalized bands from the {\it quasiparticle} Hamiltonian also have nodal lines within a few meV around the Fermi energy.

In crystals that exhibit a glide mirror plane, as is the case in \Ce{}, each band has a well defined mirror eigenvalue within the corresponding plane in reciprocal space. Bands with different eigenvalues do not interact with each other, and hence they may cross on a line in this plane instead of at an isolated point\cite{fang2016}.

\Ce{} has six glide mirror planes in which we can search for nodal lines. As we are considering a spinfull system, the glide mirror eigenvalues of the bands at momentum $\mathbf{k}$ are $\lambda = \pm i e^{i\mathbf{k}\cdot\boldsymbol{\tau}}$ where $\boldsymbol{\tau}$ is the translation vector of the glide mirror symmetry operation.\cite{young2015} These eigenvalues are smooth functions of momentum unless two bands cross. Thus by searching such discontinuities, we can identify nodal lines. To put it differently, $\lambda e^{-i\mathbf{k}\cdot\boldsymbol{\tau}}$ is constant on patches of the mirror plane that are bounded by nodal lines.

Since all glide mirror planes are conjugate to each other, i.e. symmetrically equivalent, we exemplarily show these patches of constant mirror eigenvalue in the plane perpendicular to $(1,-1,0)$ centered at $\mathbf{k}=0$ in Fig.~\ref{fig:mirror} for a quasiparticle band close to the Fermi energy. We obtain multiple nodal lines as can be seen by the many patches. The nodal lines near the center cross the Fermi energy and thus may contribute to the specific heat and spontaneous Hall effect. The ellipsoidal ones at the top and bottom are about 20 meV above; the remaining four ellipsoidal nodal lines are about 2 meV below.

To demonstrate the effect of the nodal lines on the Berry curvature, we plot the norm (or magnitude) of the latter for the same mirror plane in Fig.~\ref{fig:berry}. By comparison with Fig.~\ref{fig:mirror} we see that not all nodal lines cause a large contribution to the curvature and most of it is concentrated in the central Brillouin zone region. Conversely, there are also Berry curvature contributions of similar magnitude which do not follow the nodal lines. They show up as bright spots in the plot, but do not belong to singularities in the mirror plane.
\footnote{In principle they could correspond to Weyl nodes outside the plane, but the ones we found in the previous section are far away from these spots.}

\section{\label{sec:Conclusion}Conclusion}

We studied the Kondo semimetal \Ce{} within DFT and DMFT and observed that a Kondo resonance emerges involving only two out of the six Ce-$4f_{5/2}$ orbitals (the bonding combination of $j_x=\pm3/2$ and $\mp5/2$). Below the Kondo resonance, these are the only occupied 4$f$ orbitals, whereas at elevated temperatures all six  Ce-$4f_{5/2}$ orbitals are filled more equally. This theoretical prediction can be tested in future resonant inelastic x-ray (RIXS) and x-ray absorption experiments.

The dispersion of the  two bonding  Ce-$4f_{5/2}$ orbitals in DMFT is qualitatively different from DFT and from DFT with the 4$f$'s in the open core. It shows Weyl nodes  close to the Fermi surface. This strengthens the interpretation of \Ce{} being a Weyl-Kondo semimetal from transport measurements\cite{Dzsaber2017,Dzsaber2021}. 

Furthermore, multiple nodal lines in mirror planes provide Berry curvature near the Fermi energy that could, similarly to the Weyl points, contribute to the giant spontaneous Hall effect\cite{Dzsaber2021}. Weyl points and nodal lines are within only 3 meV of the Fermi energy at the lowest temperature reached in our calculation. There, the Kondo peak is not yet fully developed. Hence, it appears likely that the Weyl points  hit the Fermi energy when the Kondo peak further sharpens and shifts upon lowering temperature, or when marginally doping \Ce{}. 

\begin{acknowledgements}
  We thank Silke Paschen, Anna Kauch and Markus Wallerberger for very helpful discussions.
  This work has been supported by the Austrian Science Fund
(FWF) through 
  SFB Q-M\&S (FWF project ID F86) and  the Research Unit  FOR5249 `QUAST' by the Deutsche Foschungsgemeinschaft with funding obtained through  FWF project ID I~5868.
  Calculations have been mainly done on the Vienna Scientific Cluster (VSC).
  
  The DFT and DMFT data  are available from XXXX

\end{acknowledgements}

\bibliography{Ce3Bi4Pd3}

\end{document}